# Single-mode Near-infrared Lasing in a GaAsSb/GaAs Nanowire Superlattice at Room Temperature

Dingding Ren[1], Lyubomir Ahtapodov[1], Julie S. Nilsen[2], Jianfeng Yang[3], Anders Gustafsson[4], Junghwan Huh[1], Gavin J. Conibeer[3], Antonius T.J. van Helvoort[2], Bjørn-Ove Fimland[1], and Helge Weman[1,5]

[1] Department of Electronic Systems and [2] Department of Physics, Norwegian University of Science and Technology (NTNU), NO-7491 Trondheim, Norway.
[3] Australian Centre for Advanced Photovoltaics, University of New South Wales, Sydney, New South Wales 2052, Australia
[4] Solid State Physics and NanoLund, Lund University, Box 118, SE-22100 Lund, Sweden
[5] CrayoNano AS, Otto Nielsens vei 12, NO-7052 Trondheim, Norway

**Abstract:** Semiconductor nanowire lasers can produce guided coherent light emission with miniaturized geometry, bringing about new possibility for a variety of applications including nanophotonic circuits, optical sensing, and on-chip and chip-to-chip optical communications. Here, we report on the realization of single-mode room-temperature lasing from 890 nm to 990 nm utilizing a novel design of single nanowires with GaAsSb-based multiple superlattices as gain medium under optical pumping. The wavelength tunability with comprehensively enhanced lasing performance is shown to result from the unique nanowire structure with efficient gain materials, which delivers a lasing quality factor as high as 1250, a reduced lasing threshold ~ 6 kW cm$^{-2}$ and a high characteristic temperature ~ 129 K. These results present a major advancement for the design and synthesis of nanowire laser structures, which can pave the way towards future nanoscale integrated optoelectronic systems with stunning performance.

**Introduction**

Semiconductor nanowire (NW) lasers can produce nanoscale localized coherent light emission for a variety of laser applications from biomedical sciences to information technology.[1–3] The large effective refractive index of semiconductor NWs with diameters in the range of a few hundred nanometers allows for strong optical modal confinement in the radial direction, and the coherent light is extracted out directionally from few micrometers long NW Fabry–Pérot (FP) cavities, which is distinct from other types of nanolasers.[4–8] Proof-of-principle NW lasers based on bulk[1], core-shell[9] and dots-in-wire[10] gain structures have been developed to realize lasing. However, there is very limited wavelength tunability in the near-infrared demonstrated for NW lasers at room temperature, which is needed for optical interconnects in future compact nanophotonic circuits and on-chip communication systems. Thus, an alternative architecture that enables room-temperature lasing at the desired near-infrared wavelength, e.g. ~ 980 nm, is highly demanded.

To date, FP NW laser structures with ZnO[1] and ZnS[11] for ultraviolet, III-nitride[2,12] for visible, and GaSb[13], GaAs[14,15] and III-P[16] for near-infrared lasing have been demonstrated. NW lasers with binary semiconductor bulk gain are straightforward to synthesize, but achieving low lasing threshold, high lasing quality factor and wide wavelength tunability is much more challenging. This will require incorporation of controllable and complex low-dimensional heterostructures within the NW.[17] Specifically in the near-infrared range, demonstrated lasing emission from NWs with radial core-shell quantum wells are predominantly at, or close to, the bandgap of the binary core,[18–20] possibly due to the challenges in achieving a homogeneous ternary shell.[21] In current axial heterostructured NWs, e.g. quantum dots-in-wire structures, the available states for radiative recombination are restricted by both the limited number of available states in one quantum dot and the limited number of quantum dots integrated in one NW. Again, this results in lasing close to the band gap of the binary core material.[10] In addition, nanopillar-based lasers with InGaAs multi-quantum wells have only been shown to lase at low temperature.[22] Thus, an alternative NW laser architecture with both good compositional tunability and sufficient optical gain for single-mode lasing operation at room temperature is desired. Axial superlattices along the NW can provide an accumulatively long gain medium of lower bandgap materials with a sufficient number of available states for radiative recombination,[23,24] and may therefore be such an ideal NW laser structure.[25,26]

Here, we demonstrate single NW lasers with GaAsSb-based multiple superlattices as gain medium synthesized by a position-controlled Ga self-catalyzed growth method. Single-mode lasing has been observed at room temperature from 890 nm to 990 nm by adjusting the Sb content in the superlattices, giving a lasing quality factor as high as 1250 with high characteristic temperature of 129 K. In addition, the lasing threshold has also been reduced from ~ 12 kW cm$^{-2}$ to ~ 6 kW cm$^{-2}$ with higher Sb content and deeper potential wells in the superlattices.

## Results

**Growth and structural characterization of NW superlattice lasers.** Recently, self-catalyzed GaAsSb NWs have demonstrated good optical emission tunability in the near-infrared region.[27–29] In this study we have grown multiple GaAsSb-based superlattices in GaAs NWs, where such tunability is demonstrated combined with high lasing performance at room temperature (high lasing quality-factor and single-mode). The NW laser is schematically designed and shown in Fig. 1a and b before the growth and characterization experiments. Six GaAsSb superlattices were separated by GaAs spacers, and each superlattice contains ten superlattice periods of high Sb-content GaAsSb potential wells with nominal GaAs barriers in between. Three different Sb composition profiles (samples A-C with increasing Sb content from A to C) were grown and investigated. The NW superlattice structures were grown by self-catalyzed molecular beam epitaxy (MBE) on Si(111) substrates with an electron beam lithography patterned thermal oxide mask, giving a high yield of uniform vertical NWs in a regular array, as shown in Fig. 1c and d (for growth details see the Methods section). The bright-field transmission electron microscopy (BF-TEM) image in Fig. 1e shows a representative NW laser with a diameter of ~ 400 nm and a length of ~ 10 µm from sample B. To extract the compositional profile of the superlattices in the axial direction of the NW, we performed high-angle annular dark-field scanning TEM (HAADF-STEM) combined with energy dispersive x-ray (EDX) spectroscopy on a ca. 100 nm thick slice extracted from the center of a NW using focused ion beam milling. As an example, the compositional profile of Sb in the fourth superlattice, which is marked by an orange rectangle in Fig. 1e, is presented in Fig. 1f based on EDX measurements. This "sawtooth" compositional profile shows that the Sb content varies periodically with a maximum nominal content of ~ 5 at. %, which leads to periodic triangular quantum wells along the NW axis (see Supplementary Section I). The Sb-profiles for the other five superlattices in the same NW are given in Supplementary Section II for sample B. In general, the Sb content builds up in the first superlattice and a stable superlattice triangular potential is formed from the second superlattice. We emphasize here that both the high Sb solubility in the Ga catalyst and the reduction of supersaturation by the Sb incorporation in the GaAs NW lead to a reservoir effect.[27] Using a short-time supply of Sb flux, this reservoir effect results in the formation of relatively wide GaAsSb quantum wells with gradually decreased Sb content. Thereby severe radial growth induced compositional variations are avoided.

**As-grown optical characterizations.** The optical characteristics of the as-grown vertical NW array in sample B have been examined by both cathodoluminescence (CL) and micro-photoluminescence (µ-PL) spectroscopy. Figure 1g shows the low-temperature (8 K) CL spectrum, and the spatially resolved CL map for emission in the 900 to 938 nm region (red colored regions) is overlaid with the corresponding cross-section SEM image, demonstrating that most of the light emission in this region comes from the six GaAsSb-based superlattices with a dominant peak at ~ 925 nm. The other two emission peaks at ~ 875 nm and ~ 820 nm, seen in the CL spectrum in Fig. 1g, can be found to originate from the GaAs spacers between the six superlattices and the GaAs tip in the false-color image Fig. 1h, respectively. Since the dominant emission intensity is from the six GaAsSb-based superlattices, this indicates that they are optically efficient due to localized emission from inter-subband transitions in the GaAsSb quantum wells of the superlattices. In addition, we found periodic modulation of the spontaneous emissions from the superlattices, as indicated by the red dashed lines in Fig. 1g. This periodic modulation of the emission is consistently observable along the entire NW from the spatial-resolved CL spectrum in Fig. S6, indicating the presence of FP modes inside the NW (see Supplementary Section IV). The room-temperature µ-PL spectrum from the as-grown vertical NW array shows a dominant emission at ~ 955 nm, as shown in Fig. 1i, which demonstrates that the superlattices are optically bright also at room temperature.

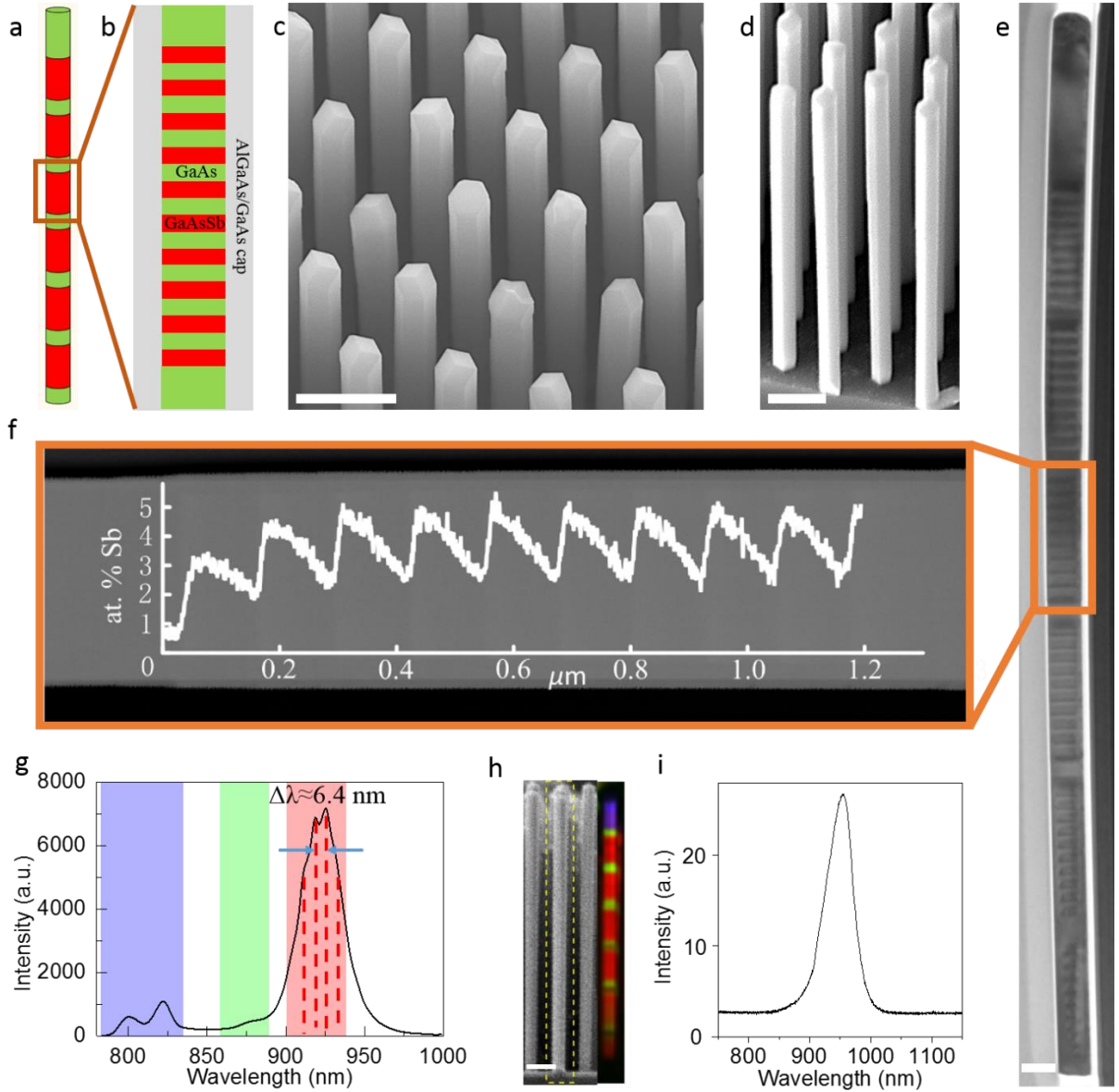

Figure 1. **Structural and optical properties of the NW superlattice laser (sample B). a,** Schematics of the NW consisting of six periodic GaAsSb-based superlattices (red) separated by GaAs spacers (green) and **b,** schematic design of each superlattice. **c, d,** 20° **(c)** and 30° **(d)** tilted-view SEM images of an as-grown NW array on a Si(111) substrate with an EBL patterned oxide mask, from the center and cleaved edge respectively. **e,** Cross-section BF-TEM image of a representative NW from **(c)**. **f,** Cross-section HAADF-STEM with the correlated EDX Sb composition profile of the fourth NW superlattice, marked by the orange rectangle in **(e)**. **g,** Average low-temperature CL spectrum of the as-grown NW array with color coding for three different spectral regions as used to construct the CL false-color image in **(h)**. The vertical red dashed lines represent the periodic peaks of the spontaneous emission originating from the FP modes of the NW laser cavity. **h,** Cross-section SEM image of the as-grown NWs and the corresponding low-temperature CL false-color image. **i,** Room-temperature μ-PL spectrum of the as-grown NW array at a pumping/excitation power density of 125 W cm$^{-2}$. The scale bars are 1 μm, 1 μm, 500 nm and 1 μm in **(c)**, **(d)**, **(e) and (h)**, respectively.

**Single NW pumping experiments.** A pulsed Ti:sapphire laser set to 800 nm was used for optical pumping experiments (see details in Methods section). Figure 2a shows a representative NW with a length of ~ 10 μm and a diameter of ~ 435 nm at the tip and ~ 395 nm at the bottom (i.e. slightly anti-tapered) from sample B. The emission spectra under pulsed excitation are shown in Fig. 2b. By increasing the excitation power density, a superlinear increase of the PL peak intensity at 950 nm is observed beyond a lasing threshold as low as ~ 6 kW cm$^{-2}$. This lasing threshold is lower than other reported NW lasers with bulk or quantum dot structures.[10,15] The Gaussian function-fitted integrated emission intensity vs. pumping power density (L-L curve), is presented in Fig. 2c on a logarithmic scale. From this L-L curve a spontaneous emission coupling factor $\beta$ ~ 0.015 is estimated.[14,22] A sudden reduction in the full width half maximum (FWHM) to ~ 0.76 nm was observed at the lasing threshold, as indicated in the Gaussian fitted function shown in the inset of Fig. 2b. The lasing quality (Q) factor, is estimated to be ~ 1250. Such small FWHM and high Q factor has not been reported previously at room temperature in III-V NW lasers using bulk[14] or quantum dot[10] as gain medium. To further confirm the lasing behavior, an optical image of the NW above the lasing threshold, is shown in Fig. 2d. The bright emission spots at both ends of the NW with a clear interference pattern are consistent with the finite-difference-time-domain (FDTD) simulated interference pattern (Fig. 2e). This is a proof of a strong wave-guiding effect in a NW FP cavity. It should be noticed that emission through the NW sidewalls quite weak compared to the lasing from the end facets, indicating a strong radial optical confinement, with a relatively small proportion of waveguide leakage from the NW cavity. Figure 2f shows the normalized lasing spectra above threshold at various temperatures from room temperature to 10 K. The decrease of the lattice temperature results in a blue shift of the NW lasing peak from ~ 950 nm to ~ 883 nm, which is well below the bandgap of GaAs (marked as vertical black dashed lines) in the whole temperature range, showing that the lasing occurs in the GaAsSb-based NW superlattices. A characteristic temperature as high as ~ 129 K is attained by fitting the exponential function of the lasing threshold power vs. temperature to the data points in Fig. 2g. This is among the highest values of characteristic temperatures for III-V NW lasers reported so far.[10,15] We note that optical pumping at 10 K gives a laser wavelength shift towards higher energy with increasing excitation power (see Supplementary Section VI), which is attributed to band filling effects.

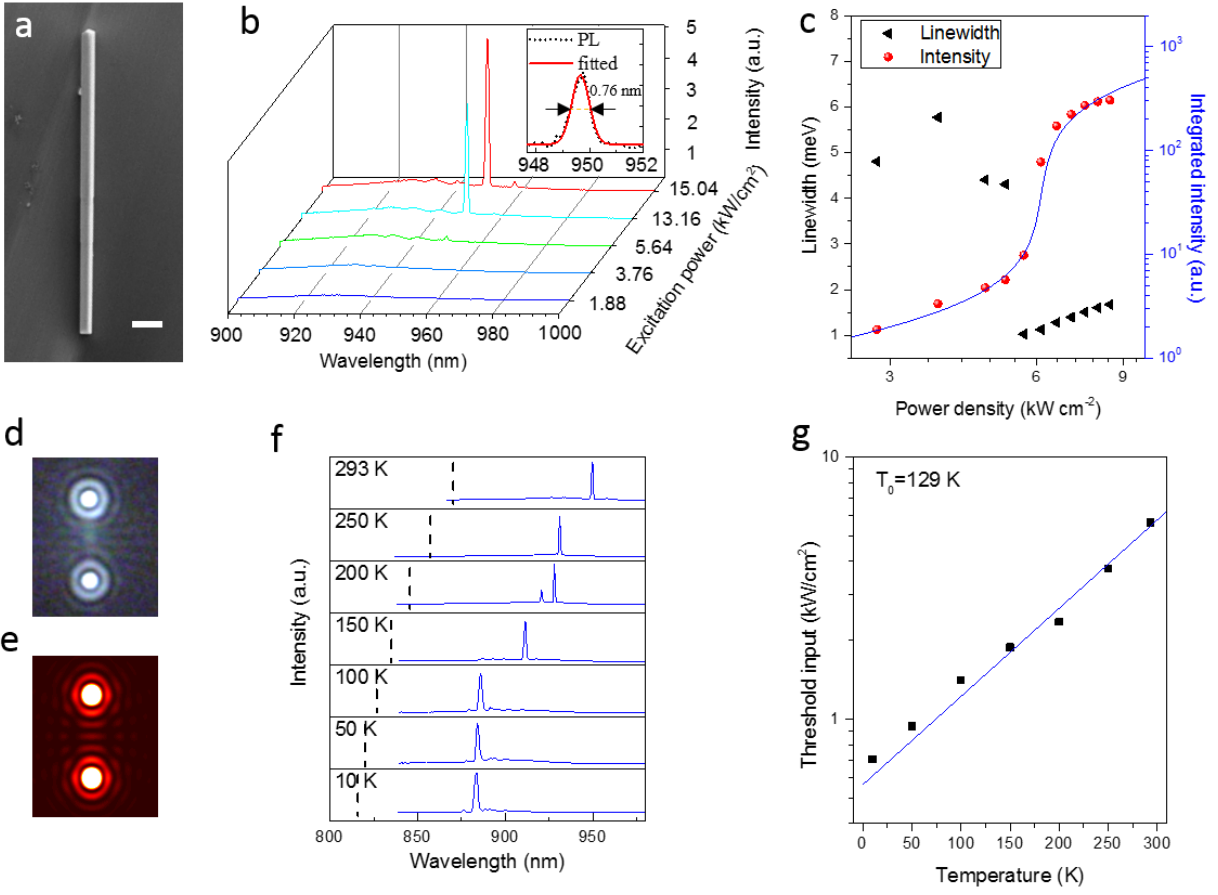

Figure 2. **Lasing properties from a single NW superlattice laser (sample B). a**, SEM image of a single NW superlattice laser on a SiO$_2$/Si substrate (scale bar 1 μm). **b**, Emission spectra for a single NW laser under pulsed excitation at different power densities. The inset shows that the laser peak is well fitted with a Gaussian function with a FWHM ~ 0.76 nm. **c,** Double logarithmic L-L curve of integrated emission intensity and corresponding linewidth of the lasing peak vs. optical pumping power density. **d**, Experimentally observed interference pattern from the NW laser above lasing threshold. **e**, Corresponding FDTD-simulated interference pattern to **(d)**. **f**, Temperature-dependent lasing emission spectra (normalized intensity) at temperatures from 10 K to 293 K. The vertical black dashed lines represent the wavelength corresponding to the GaAs bandgap, showing that the lasing from the GaAsSb-based superlattice is well below the GaAs bandgap energy. **g,** Plot of laser threshold pump power vs. temperature, indicating a fitted characteristic temperature of ~ 129 K.

**Polarization measurements and simulations.** To further evaluate the lasing properties from the NW superlattice laser, both linear polarization emission measurements and FDTD simulations (see Supplementary Section III) have been performed. The intensity of the polarized PL emission from the NW as a function of the angle to the NW growth axis, is shown in Fig. 3a. The degree of linear polarization, $(I_\perp - I_\parallel)/(I_\perp + I_\parallel)$, was determined to be 87 % for the NW laser shown in Fig. 2, based on a least-square fit to the experimental data. The lasing mode is thus strongly polarized perpendicularly to the NW axis (and the lineshape is Gaussian), indicating single-mode lasing. FDTD simulations indicate that this is most likely the $HE11_b$ NW waveguide mode since this mode is strongly confined to the center part of the NW and also has the highest effective refractive index (see Supplementary Section III), as shown in Fig. 3b. It thus has a large overlap with the active GaAsSb-based NW superlattice gain medium.

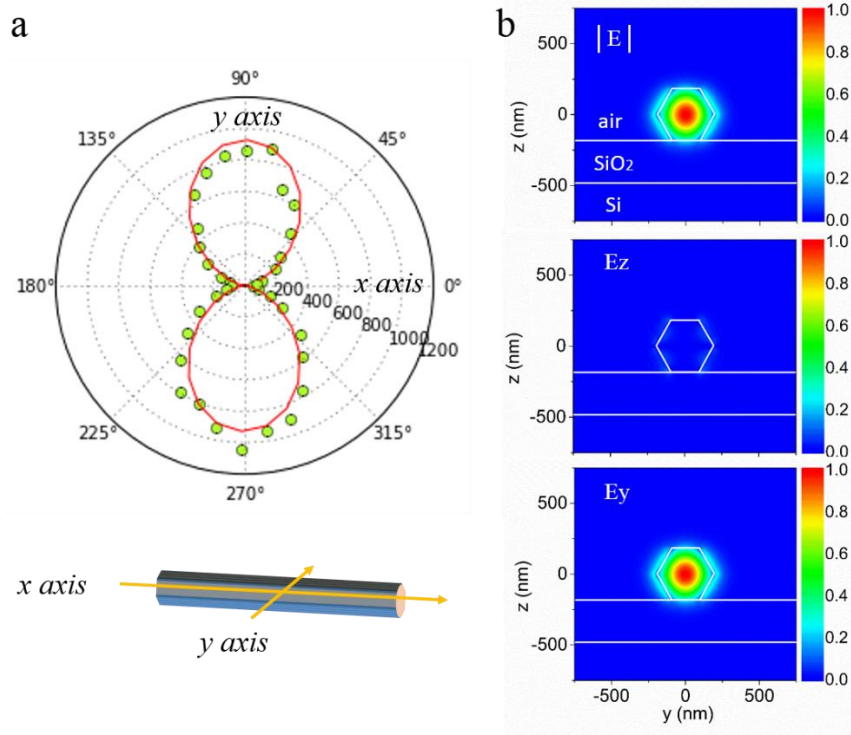

Figure 3. **Polarization measurement and cross-section view of the optical waveguide mode profile of the NW superlattice laser (the NW is shown in Fig. 2a from sample B). a,** Polar plot of the linear polarization intensity of the laser emission. The polarization angle is measured in the x-y plane of the NW, parallel to the substrate. **b,** FDTD simulations of the Normalized electric field profile of the $HE11_b$ NW waveguide mode, showing that this mode is strongly polarized in the y-direction and parallel to the x-y substrate plane. Red and blue colors represent the highest and lowest electric field intensity of the $HE11_b$ NW waveguide mode, respectively.

**Wavelength tuning of the lasing emissions.** The wavelength tunability of the NW laser is an important motivation to explore the GaAsSb-based superlattice design and an important prerequisite for practical applications in the near-infrared wavelength region. By tuning the Sb compositional profile in the superlattices, either by the Sb flux or the GaAsSb growth time in the superlattices, the wavelength emission of the NW laser can be tuned from 890 nm to 990 nm, as shown in Fig. 4a. With a reduction of the Sb content by reducing the Sb flux, the lasing wavelength blue-shifts, leading to lasing emissions around 890 nm with an Sb content of ~ 1 at. % in sample A as measured by EDX, shown in Fig. 4b. With higher Sb content in the GaAsSb superlattice, the lasing emission red-shifts. The longest wavelength lasing emission observed is at 990 nm (see Fig. 4d), which comes from the superlattice with the highest Sb content (sample C). ~ 8.0 at. % Sb was measured by EDX in sample C and it was obtained by doubling the GaAsSb potential well growth time as compared to sample B. It should be noticed that for the NW superlattice structure with the lowest Sb content, the lasing is achieved at a much higher laser threshold (~ 12 kW cm$^{-2}$) compared to the NW lasers with higher Sb content. (see Supplementary Section VIII) This indicates that the room-temperature lasing threshold strongly depends on the Sb profile of the GaAsSb-based potential wells in the NW superlattices.[17]

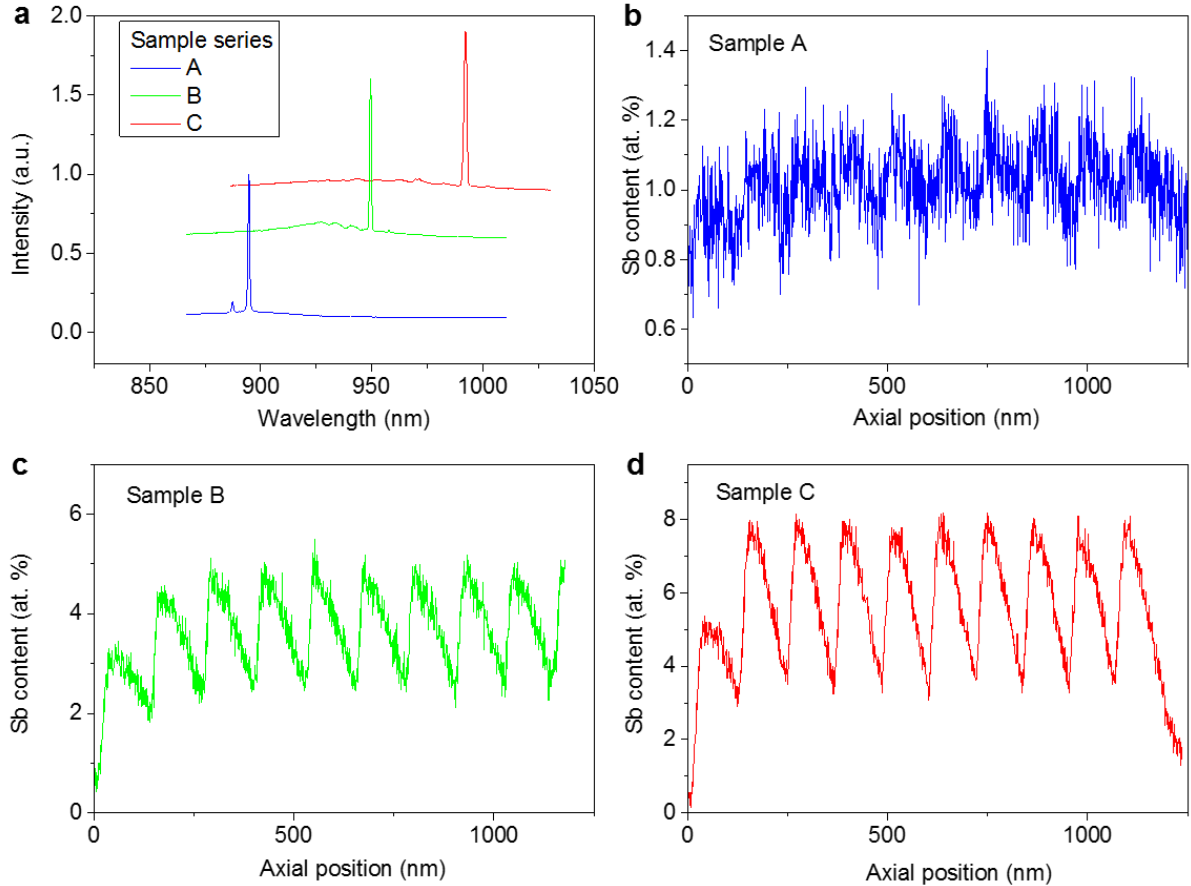

Figure 4. **Laser emission spectra and Sb content profiles (at. %) for NW superlattice Samples A, B, and C. a,** Laser emission tuned from around 890 nm (Sample A), 950 nm (sample B) to 990 nm (sample C) by controlling the Sb composition in the GaAsSb-based superlattices. **b-d,** Corresponding EDX Sb composition profiles from the 4$^{th}$ GaAsSb-based superlattice of Samples A, B and C, respectively.

**Discussions**

To better understand the lasing mechanism from the NW superlattice lasers, gain simulations have been performed based on the compositional profiles of Sb in Fig. 1f (see Supplementary Section I and VII). The electrons in the conduction band are shared among all GaAsSb quantum wells in one superlattice due to the shallow conduction band potential, while the holes are confined in each GaAsSb quantum well in the valence band, at least for the highest-lying subbands. The length of the gain medium is estimated to be ~ 30 nm per GaAsSb-based quantum well in the superlattices, based on the overlap between the electron and hole wave functions in the fundamental subbands in the conduction and valence band, respectively. Due to the strong hole confinement, different GaAsSb-based quantum wells will contribute to the active medium gain independently. (For further explanations see Supplementary Section VII). This correlates well with the Gaussian shape of the emission peak at the lasing threshold, which results from the electronic difference of the GaAsSb-based multiple quantum wells (see Supplementary Section V). The total material gain at lasing threshold is estimated to be ~ $4.1 \times 10^3$ cm$^{-1}$ at room temperature from FDTD and material gain simulations. Although this room-temperature value is ~ 2.25 time higher than previously reported bulk-gain GaAs NW lasers[14,30], the length of the gain region is only one quarter of the bulk-gain NW laser. Compared to the bulk-gain NW structure, our NW superlattice structures have a much smaller gain volume without sacrificing the carrier collection efficiency, which significantly increases the carrier density in the active gain medium under the same optical pumping power.[31] This makes the population inversion in GaAsSb-based NW quantum wells easier than in a bulk-gain NW structure, and the lasing threshold is thus reduced. Moreover, our NW superlattice structure has a highly-repeatable well-controlled compositional profile and high radiative recombination rates, preventing severe band filling-induced blue-shifts below the lasing threshold and enabling lasings at longer near infrared wavelengths with minimized re-absorption loss. Last but not the least, the gain media are located at the core of the NW, which only supports the fundamental HE11$_b$ waveguide mode, bringing about high optical confinement factor and single-mode lasing.

In conclusion, we have presented the excellent lasing properties from GaAsSb-based NW superlattices. By adjusting the Sb-compositional profiles in the NW superlattices, the single-mode laser emission wavelengths can be tuned from 890 nm to 990 nm at room temperature. Due to its unique superlattice design with efficient carrier collection and radiative recombination, the NW superlattices deliver unprecedented sub-nanometer linewidth lasing with a high lasing Q factor (~ 1250) and high characteristic temperature (~129 K) with reduced lasing threshold. The NW superlattice structure thus provides a new concept in designing tunable and efficient NW laser sources, promising as useful building blocks for future nanophotonic light sources and optical interconnects.

**Methods**

**NW growth:** The GaAsSb-based NW superlattices were grown at a substrate temperature of 625 °C using a solid source MBE (Veeco GEN 930). The Ga and $As_2$ fluxes used in this study were 0.7 ML/s and $2.5 \times 10^{-6}$ Torr, respectively. The growth of the NW array with superlattices were initiated by suppling a Ga flux for 45 s to form the Ga catalyst in the holes of the EBL-patterned oxide mask on a Si(111) substrate, and then a short (~ 100 nm) NW stem of GaAsSb were grown for 1 min with an $Sb_2$ flux of $1 \times 10^{-7}$ Torr to improve the vertical yield in the array. Subsequently, a GaAs NW segment was grown for 2 min before the growth of the GaAsSb-based NW superlattice. Each GaAsSb superlattice period in the six superlattices was grown by first keeping the Ga shutter closed and the Sb and As shutters open for 10 s (samples A and B) or 20 s (sample C), followed by 96 s with Sb shutter closed and Ga and As shutters open. The GaAs spacers between each superlattice were grown with Ga and As shutters open for 3 min. $Sb_2$ flux during superlattice growth was $0.3 \times 10^{-6}$ Torr (sample A) or $1 \times 10^{-6}$ Torr (samples B and C). The Ga catalyst solidification process was performed by supplying an $As_2$ flux of $1 \times 10^{-5}$ Torr for 15 min, followed by the growth of an $Al_{0.33}Ga_{0.67}As$ shell (14 nm) and a GaAs cap (6 nm) to passivate NW surface states.

**Electron microscopy:** For the TEM characterization, the NW superlattice lasers were first transferred from the substrate to a 50 nm thick SiN window TEM grid using a diamond scraper. The NWs were characterized by conventional TEM using a JEOL JEM-2100F at 200 kV. For each of the three samples in this study, one structurally representative NW laser was selected to be thinned by focused ion beam (FIB), using a FEI Helios NanoLab DualBeam, for a more detailed TEM study. The prepared TEM lamellas were ~ 12 µm long and ~ 100 nm thick and included the entire NW laser. The specimens were characterized with HAADF STEM and EDX using a JEOL JEM ARM200F at 200 kV with a Centurio SDD EDX system (solid angle 0.98 sr). EDX data were acquired with the NW laser specimen on the [112] zone-axis. For the quantitative analysis, the EDX data were first denoised using principal component analysis before a Cliff Lorimer based quantification routine was applied with calculated k-factors. All data analysis was performed using the Python based library HyperSpy.

**Optical spectroscopy:** For the optical pumping experiments, the as-grown NW superlattice lasers were mechanically detached from the Si substrate using a sharp diamond-tip scriber and transferred to a $SiO_2$/Si substrate with an oxide thickness of 300 nm. The optical pumping was achieved with a mode-locked Spectra Physics Tsunami Ti:Sapphire laser delivering 300 fs pulses at 800 nm. Optical excitation and collection were done with a Mitutoyo 50× NA0.65 infinity corrected microscope objective (50 % transmission loss at the excitation wavelength). The beam spot was defocused to ~ 8 µm in order to excite most of the NW superlattices. The PL emission was dispersed with a Horiba Jobin-Yvon iHR500 spectrometer and detected with an Andor Newton EMCCD camera. For the low temperature experiments, a Janis ST-500 optical cryostat was used. The CL was done in hyperspectral mode in a dedicated SEM with a cold stage (Liquid He, 8 K). The data was recorded with software from DELMIC, using an Andor Newton CCD, connected to a SPEX monochromator.

**Author contributions**

D.R. conceived the idea and designed the structures under supervision from B.O.F. and H.W.. D.R. carried out the growth experiments, SEM and preliminary TEM characterizations. D.R. and L.A. performed the PL and optical pumping measurements. L.A. performed the electronic structure and gain simulations in cooperation with D.R.. J.S.N. carried out the TEM characterizations under supervision from A.T.J.v.H.. J.Y. performed the FDTD simulations under supervision from G.C.. A.G carried out the CL measurements. J.H. contributed with data analysis. D.R. wrote the manuscript with contributions from all authors. B.O.F. and H.W. supervised the project.

**Acknowledgements**

We would like to thank Tron Arne Nilsen and Dong-Chul Kim for their kind technical supports. This work was supported by the FRINATEK (Grant 214235) and NANO2021 (Grant 239206) programs of the Research Council of Norway. The Research Council of Norway is also acknowledged for the support to NTNU NanoLab through the Norwegian Micro- and Nano-Fabrication Facility, NorFab (197411), the NORTEM facility (197405) and the Norwegian PhD Network on Nanotechnology for Microsystems (FORSKERSKOLER-221860). We also acknowledge the support from the Swedish Research Council (VR), the Foundation for Strategic Research (SSF), the Knut and Alice Wallenberg Foundation (KAW) and NanoLund. Jianfeng Yang would like to acknowledge the financial support from the China Scholarship Council (CSC, No. 201306070023).

# Single-mode Near-infrared Lasing in a GaAsSb/GaAs Nanowire Superlattice at Room Temperature

Dingding Ren[1], Lyubomir Ahtapodov[1], Julie S. Nilsen[2], Jianfeng Yang[3], Anders Gustafsson[4], Junghwan Huh[1], Gavin J. Conibeer[3], Antonius T.J. van Helvoort[2], Bjørn-Ove Fimland[1], and Helge Weman[1,5]

[1] Department of Electronic Systems and [2] Department of Physics, Norwegian University of Science and Technology (NTNU), NO-7491 Trondheim, Norway.
[3] Australian Centre for Advanced Photovoltaics, University of New South Wales, Sydney, New South Wales 2052, Australia
[4] Solid State Physics and NanoLund, Lund University, Box 118, SE-22100 Lund, Sweden
[5] CrayoNano AS, Otto Nielsens vei 12, NO-7052 Trondheim, Norway

# I. Lasing mechanism

To model the GaAsSb-based superlattice potential, corresponding conduction and valence band potentials were created. For the conduction band potential, the "sawtooth" Sb content variation depicted in Fig. 1f was converted to an energy band offset variation with respect to GaAs following Fig. 3 in Ref. 1. The valence band potential was constructed by calculating the GaAsSb band gap at each point of the model structure and subtracting it from the conduction band minimum potential. The electronic structure in the conduction and valence bands was calculated by solving the time-independent Schrödinger equation numerically using the Shooting method.[2] To simplify the calculation, two model structures were considered: a single, triangular GaAsSb composition profile which corresponds to a nearly triangular finite quantum well, and two such GaAsSb composition profiles next to each other which corresponds to a double finite, nearly triangular quantum well.

As has been shown previously[3,4], EDX analysis has a limited accuracy on the quantitative information about the Sb content in NWs. Thus, in the present model it was only assumed that the recorded EDX counts are proportional to the true Sb content. The bound states for electrons and holes were thus computed for a series of Sb contents in the range 1.5 – 4.5 at. %, using the Sb composition gradient calculated from the EDX data in Fig. 1f. The resulting emission energies for optical recombination between electrons in the conduction band and holes in the valence band in their ground states were then calculated using Formula 6 in Ref. 5, whereby quantum confinement was taken into account by adding the electron and hole confinement energies to the calculated bulk GaAsSb band gap. The resulting emission energies were plotted as a function of the Sb content. By comparing this calculated emission energy to the spontaneous emission PL peak for sample B (955 nm/1.298 eV (Fig. 1i)), it was determined that the actual maximum Sb content within the GaAsSb quantum well is 3.3 at. %, and that the Sb content decreases to 2.1 at. % at the upper edge of the GaAsSb quantum well.

The quantum well in the conduction band corresponding to the above described compositional profile of Sb gives rise to three bound states for electrons. In view of the one-dimensional nature of the quantum confinement in the model structure, each of these bound states is in fact the bottom of a subband. As demonstrated in Fig. S1, the consideration of the double quantum well case reveals that all of these subbands couple electronically with the neighboring quantum wells. Thus, the electrons in the conduction band will be shared among all GaAsSb quantum wells of the superlattice.

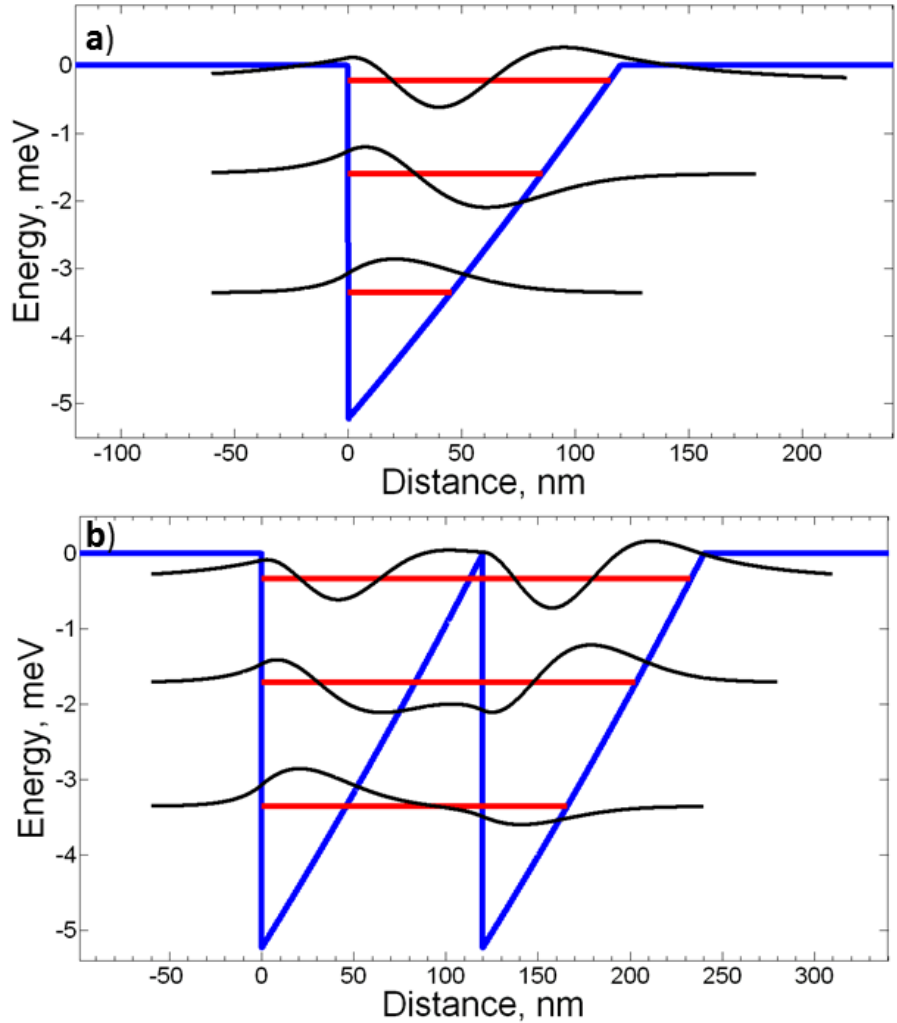

Fig. S1. (a) Conduction band potential of the single GaAsSb-based quantum well model structure (blue) together with quantized electron subband levels (red) and corresponding wavefunctions (black). (b) Conduction band potential for the double GaAsSb-based quantum well structure together with quantized electron subband levels (red) and corresponding wavefunctions (black).

The GaAsSb-based quantum well in the valence band is much deeper and contains twenty quantized hole subbands. By plotting the wave functions for the highest- and lowest-lying energy subbands, calculated in the framework of the single-quantum well model, it can be seen that, while the lowest hole subband will couple electronically with the neighbouring GaAsSb quantum wells (as was the case in the conduction band), the wave function associated to the highest subband decays much faster away from the quantum well. Thus, holes in the fundamental subband (ground state) will remain confined to the individual GaAsSb quantum wells. This means also that despite the fact that we have an electronic superlattice in the conduction band, the energy structure in the valence band is essentially still that of multi-quantum wells.

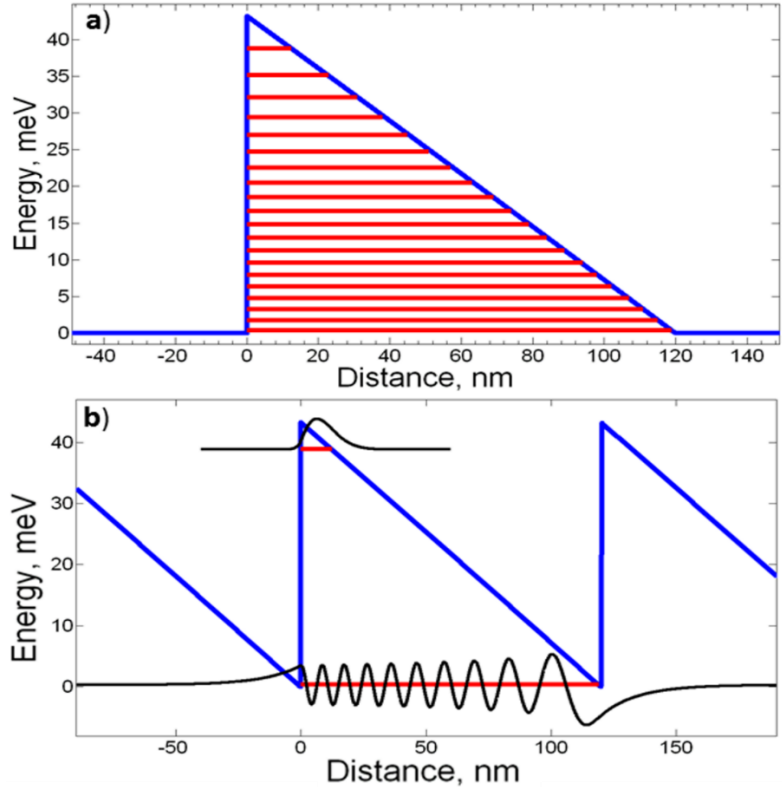

Fig. S2. (a) Valence band potential of the GaAsSb-based single-quantum well model structure (blue) together with the quantized hole subband levels (red). (b) GaAsSb-based multi-quantum-well potential (blue) together with the lowest and highest subbands (red) and the corresponding single quantum well wave functions (black).

In order to estimate the carrier occupancy of the subbands in the GaAsSb-based superlattice, we use the following parameters for the optical excitation beam: average power of 5 mW, wavelength of 800 nm, spot size defocused to 8 µm, pulse width of 300 fs, and repetition rate of 80 MHz. Since the NW diameter is ~ 400 nm and the length of one GaAsSb-based superlattice is ~ 1.2 µm, the above parameters mean that within one excitation pulse, ~ between $2.7 \times 10^3$ and $4.1 \times 10^3$ photons impinge on each superlattice, depending on whether only four (worst case scenario), five or all six (best case scenario) superlattices are excited. This corresponds to a carrier density of 1.8 - 2.7 $\times$ $10^{16}$ cm$^{-3}$ across the entire superlattice. In turn, this gives a value of $2.1 \times 10^{11}$ - $3.3 \times 10^{11}$ cm$^{-2}$ for the sheet carrier density within each insert, assuming all inserts are equally populated with carriers.

Meanwhile, the carrier sheet density inside the well, assuming that only one subband is populated, is given by [6]:

$$n_s = \frac{m^*}{\pi \hbar^2} k_B T \ln\left(1 + \exp\left(\frac{E_F - E_1}{k_B T}\right)\right),$$  (1)

where $m^*$ is the effective mass of the carrier, $T$ is the sample temperature, $E_1$ is the bottom of the first subband, and $E_F$ is the Fermi level. By choosing the Fermi level to be equal to the bottom of the second subband in the conduction band, we can estimate that at room temperature (T = 295 K), the maximum sheet density corresponding to only one occupied subband in the conduction band is $5.1 \times 10^{15}$ cm$^{-2}$. The equivalent sheet density for the first subband in the valence band is an order of magnitude higher, $4.2 \times 10^{16}$ cm$^{-2}$. Thus, the majority of the electrons and holes generated by the supplied optical excitation will occupy the states in the first subbands.

## II. Sb composition profiles in the NW superlattices

The quantitative EDX characterization of the six GaAsSb-based superlattices in NW samples A, B and C, with lasing around 890 nm, 950 nm, and 990 nm, are shown in Fig. S3**a** to **c**, respectively. For the EDX characterization, ~ 100 nm thick slices were cut out along the center of the NWs using FIB milling. Although many effects can cause the minor Sb compositional variations in the first GaAsSb-based superlattice in each sample, such as Ga-catalyst contact angle influences[7] and re-adsorption effects[8], the Sb composition profiles are quite stable from the second superlattice in each sample. These stable Sb compositional profiles lead to uniform energy levels inside the GaAsSb-based quantum wells as discussed in section I, crucial in order to increase the optical gain and reduce the NW lasing thresholds, as shown in section IV.

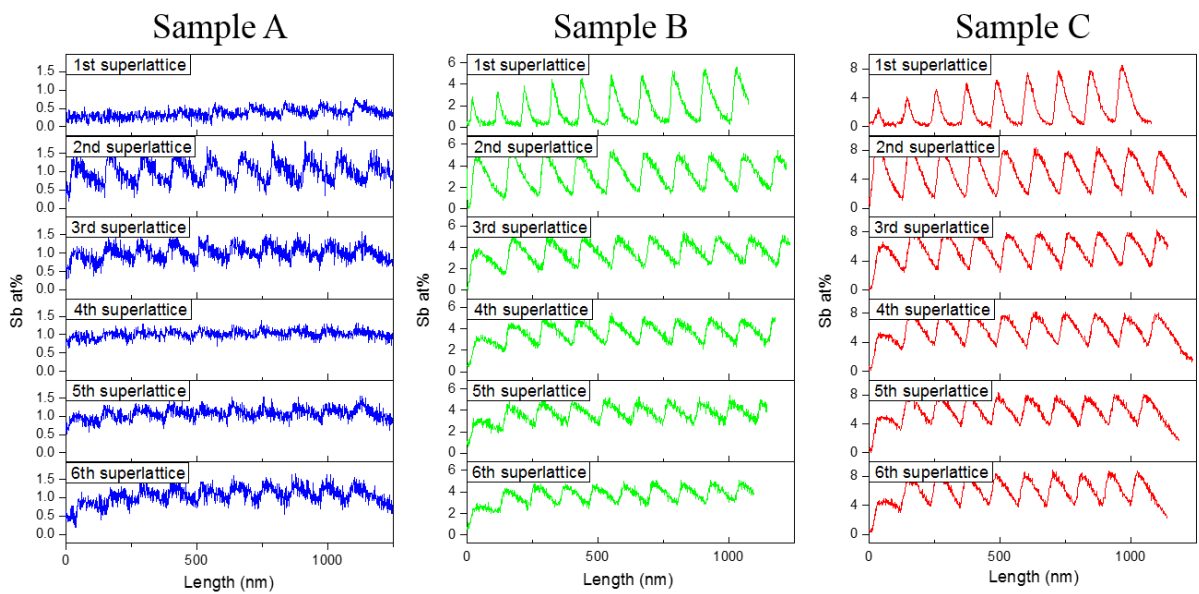

Fig. S3. Quantitative Sb EDX profiles of the six GaAsSb-based superlattices in the NWs lasing at (a) ~ 890 nm (sample A), (b) ~ 950 nm (sample B) and (c) ~ 990 nm (sample C). The plots show the Sb compositional profiles of the first to the sixth superlattice, in sequence of the growth direction from the bottom to the top of the NW.

## III. Guided mode simulation of the NW laser cavity

To analyze the lasing mode and the optical mode properties of the NW laser cavity, three-dimensional finite-difference-time-domain (FDTD) simulation was carried out using Lumerical FDTD Solutions.[9] A hexagonal geometry for the GaAsSb-based NW superlattice was used to represent the NW cavity with a constant diagonal (edge-to-edge) NW diameter of 425 nm. In the simulations the NW is modeled as lying on a $SiO_2$/Si substrate with 300 nm thick $SiO_2$. The refractive indices at 950 nm used for GaAs, $SiO_2$ and Si are 3.55, 1.45, and 3.58, respectively.

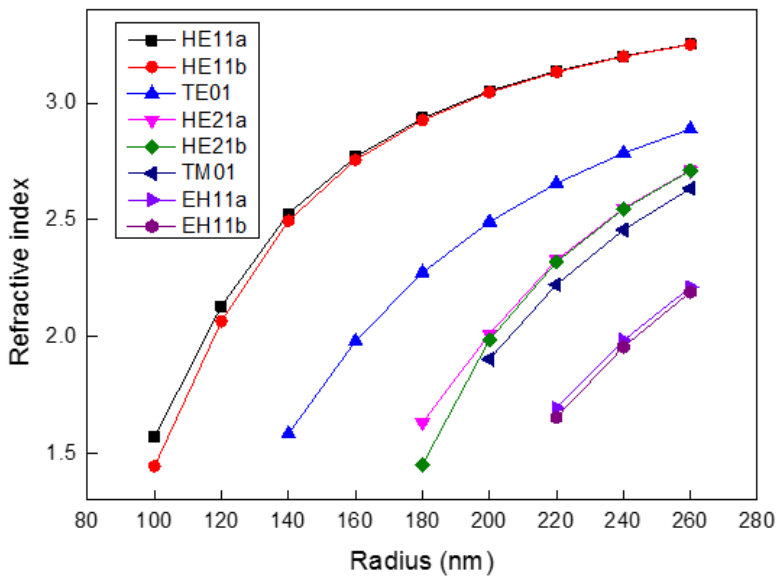

Fig. S4. Effective refractive indices of the guided modes in a NW superlattice laser cavity, placed on $SiO_2$/Si substrate, as a function of the NW radius for lasing wavelength of 950 nm.

As seen in Fig. S4, the hybrid HE11 waveguide modes have the highest effective index, which should result in a strong optical confinement of the light inside the NW cavity. The normalized electric field intensity profiles of the cross-section of the NW cavity are shown in Fig. S5, using the same NW geometry. The electrical field of the HE11 modes are confined at the core part of the NW cavity, which is different from the other modes. Considering also the AlGaAs/GaAs shell passivation of the NW core, the HE11 modes have much larger overlap with the GaAsSb-based superlattice gain medium compared to other modes. Combined with the experimental polarization measurement in Fig. 3a, it's concluded that the lasing mode is $HE11_b$, whose light electric field polarization is perpendicular to the NW axis and parallel to the plane of the substrate.

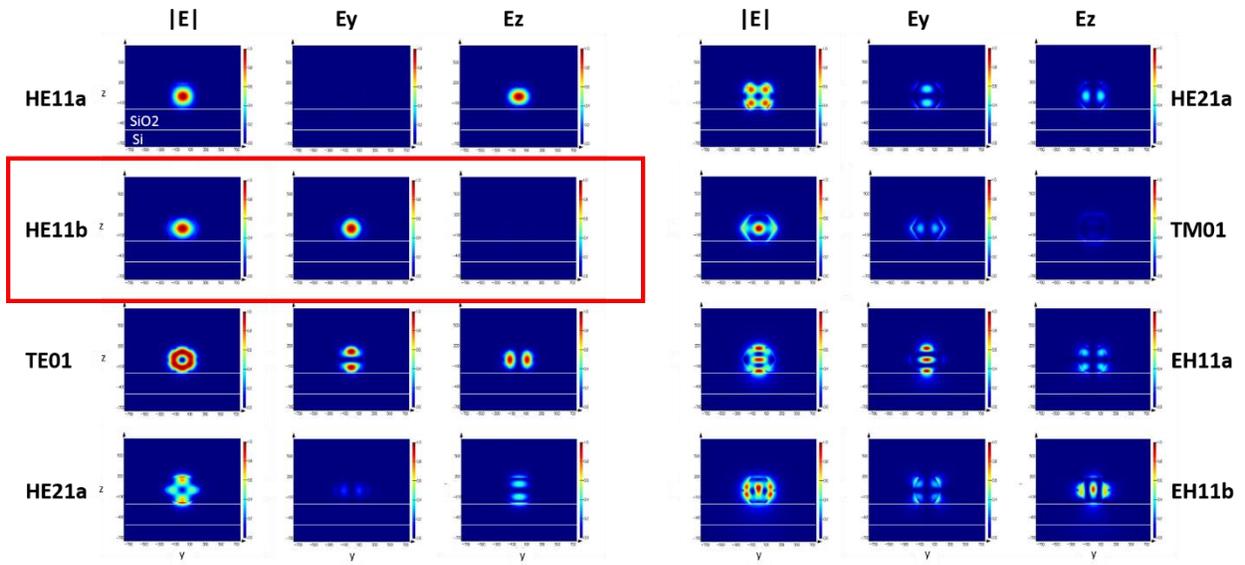

Fig. S5. NW cross-section profiles of normalized electrical field intensity from possible guided modes inside the NW superlattice laser cavity.

## IV. Direct observation of Fabry–Pérot modes from CL.

As is shown in Fig. 1h, the NW (in yellow dashed rectangle) used for the low-temperature CL measurements of sample B is as-grown and located at the cleaved edge of the sample. As this NW is partly detached away from the Si substrate and suspended in air, it forms a bottom NW/air mirror. To confirm that the periodic CL intensity fluctuations, marked by vertical red dashes in Fig. 1g, is from Fabry–Pérot (FP) modes, a spatial-resolved CL mapping is shown in Fig. S6. The vertical red arrows in Fig. S6 clearly show that the wavelength periodic intensity fluctuations are observed along the entire NW, consistent with optical confinement inside the whole NW cavity.

The group index of the HE11$_b$ mode, $n_{group}$, is calculated to be 4.94 at ~ 920 nm based on FDTD simulation. From this the energy spacing of the FP modes can be estimated by $\delta E = (\hbar \pi c)/(n_{group} \cdot L)$ [10], where $L$ is the NW length and $c$ is the speed of light in vacuum. Using $L$ = 10 μm, the spacing of the FP modes should be ~ 12.6 meV, which is in close agreement with the spacings observed in Fig. 1g and Fig. S6 (~ 10 meV).

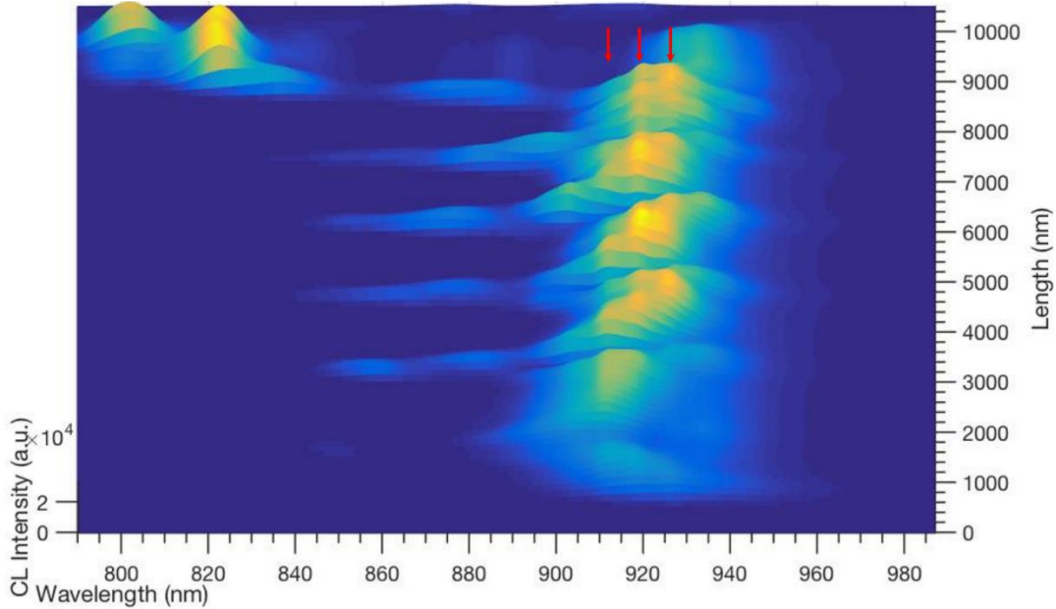

Fig. S6. Spatial-resolved low-temperature (10 K) CL mapping of emission wavelength vs NW axial position in sample B (CL measured from the NW in the yellow rectangle in Fig. 1h). The bottom of as-grown NW (NW/substrate interface) is located at ~ 500 nm and the tip of NW at ~ 10 μm.

# V. Lineshape fitting of optical emission

Theoretically, the laser emission lineshape should be a Lorentzian or a hyperbolic secant. These functions are shown below.

Lorentzian: $$\frac{\gamma^2}{(\omega_k - \nu)^2 + \gamma^2}$$

Hyperbolic secant: $$\frac{1}{\cosh\left(\frac{\omega_k - \nu}{\gamma}\right)}$$

(here $\omega_k$ is the peak laser frequency, $\nu$ is the frequency and $\gamma$ is the linewidth)

However, none of these two lineshape functions were found to fit the data properly, as shown in Fig. S7a below of NW shown in Fig. 2 from sample B.

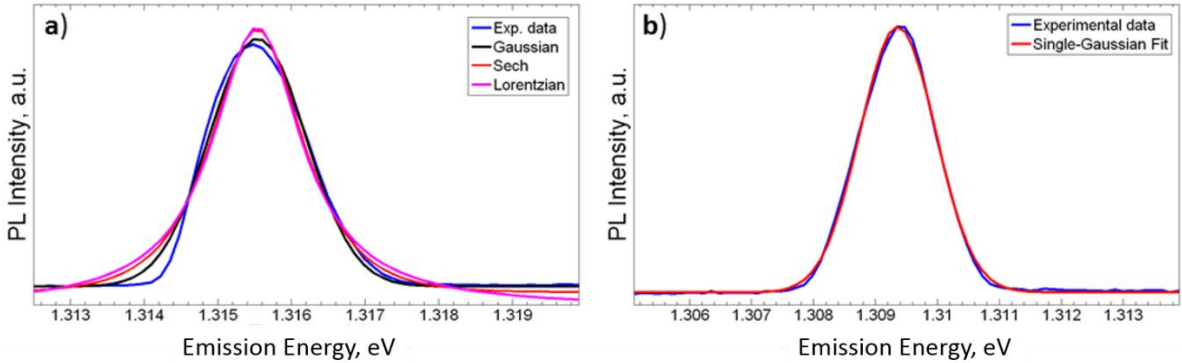

Fig. S7. (a) Gaussian, Hyperbolic secant and Lorentzian lineshape fits to a PL spectrum from NW laser shown in Fig. 2 from sample B. The optical power density was 7.52 kW cm$^{-2}$, which is above the threshold for this NW. (b) Gaussian fit to a PL spectrum from the polarization-resolved experiment from the same NW shown in Fig. 2 from sample B at 5.64 kW cm$^{-2}$, which is just above the laser threshold.

There is a noticeable discrepancy in the tails of the fit to the emission spectrum recorded for an excitation power density of 7.52 kW cm$^{-2}$ (Fig. S7a). Although this discrepancy is there for all lineshape curves, it is least pronounced for the Gaussian, compared to the Lorentzian and Hyperbolic secant fits in Fig. S7a. In this case the data can be described as "nearly Gaussian". On the other hand, the PL in Fig. S7b spectrum taken at lower power, and thus nearer lasing threshold, is essentially Gaussian.

Due to the fact that the NW superlattice laser has different quantum wells lasing independently at the same time, there is a large carrier density fluctuation throughout the active medium which causes a fluctuation of the cavity resonant frequency. In this case, a Gaussian lineshape function is expected just above lasing threshold.[11] This is also in support of the multiple-quantum-well explanation of the lasing mechanism in SI Section I.

## VI. Low-temperature optical pumping experiments

The NW laser shown in Fig. 2 in the main text were also evaluated at 10 K under pulsed excitation. The power-dependent spectra in Fig. S8 shows a single-mode lasing threshold ~ an excitation power density of 0.94 kW cm$^{-2}$ at 888 nm. By increasing the excitation power to 1.88 kW cm$^{-2}$, a second peak at 881 nm arises. At an excitation power of 3.76 kW cm$^{-2}$, the 881 nm peak becomes dominant, and the intensity of the 888 nm peak disappears in the background.

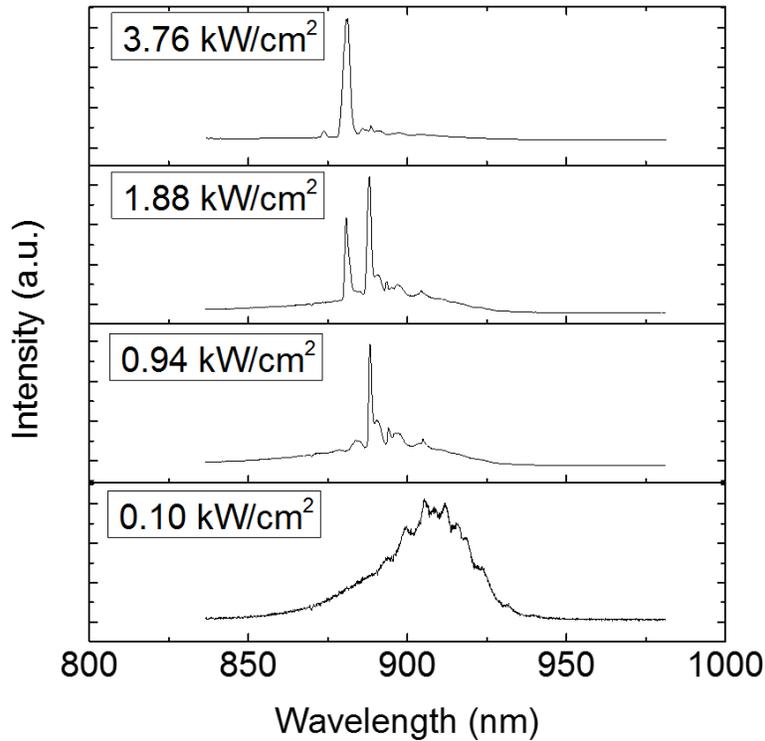

Fig. S8. Low-temperature (10K) power-dependent emission spectra of NW laser shown in Fig. 2 from sample B.

# VII. Mode confinement factor and modal gain required for lasing

The modal confinement factor $\Gamma_{core}$ has been calculated for the NW core area for each guided mode, which is given by[12]

$$\Gamma_{core} = \frac{2\varepsilon_0 c n_b \iint_{core} dxdy |\mathbf{E}|^2}{\iint dxdy [\mathbf{E}\times\mathbf{H}+\mathbf{H}\times\mathbf{E}]\cdot\hat{z}} \quad , \tag{1}$$

where $\varepsilon_0$ is the vacuum permittivity, $c$ is the speed of light in vacuum, and $\mathbf{E}$ and $\mathbf{H}$ are the complex electric and magnetic fields of the $HE11_a$ guided mode, respectively. $n_b$ is the refractive index of the gain medium at lasing wavelength. $\Gamma_{core}$ has been calculated to be 1.12 for the $HE11_b$ mode from 2D finite difference eigenmode simulations.

The wave functions for electrons and holes inside the subbands calculated in the framework of the single-quantum well model, normalized to yield unity when their absolute square is integrated from practical negative infinity to practical infinity, are plotted in Fig. S9.

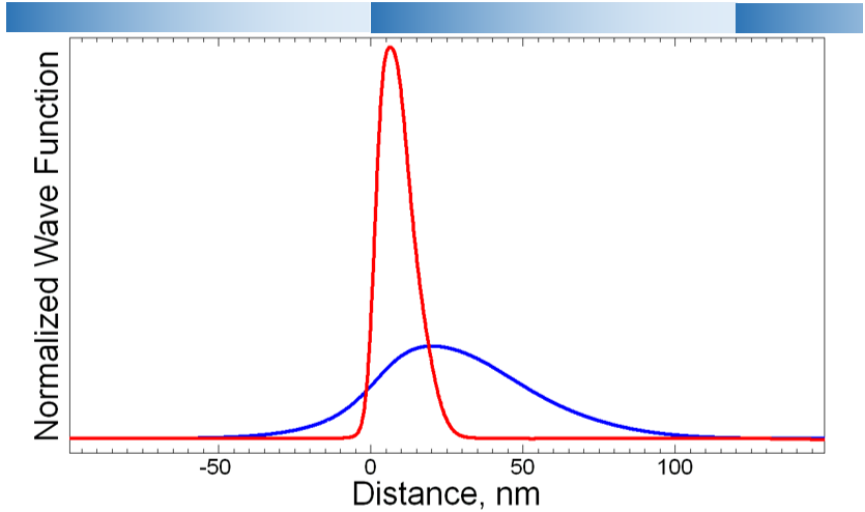

Fig. S9. Normalized electron (blue) and hole (red) wave functions for the ground state energy subbands in the GaAsSb-based NW triangular quantum well of the superlattice. The corresponding gradient of Sb content in triangle quantum wells is represented as dark blue on top of the wavefunction.

It is evident that the electron wave function is more spread out and has a lower peak value compared to the hole wave function, which is fairly concentrated inside the triangular quantum well. This is indeed expected in view of the fact that the triangular quantum well in the GaAsSb conduction band is shallower and the electron effective mass lighter compared to the case of holes in the GaAsSb valence band. The radiative recombination rate between electrons and holes is proportional to the absolute square of the overlap integral between the electron and hole wave functions given by:[13]

$$F_{cv} = \int_a^b \psi_c(z)\psi_v(z)dz \qquad (2)$$

In order to calculate the overlap integral (2), the electron and hole wave functions were fitted with a five-Gaussian envelope function. Then the width of the integration interval ($a - b$) was varied in the range from 1 to 300 nm and an optimization of $|Fcv|^2$ as a function of the lower boundary $a$ was performed for each value of the integration interval. The result is plotted in Fig. S10.

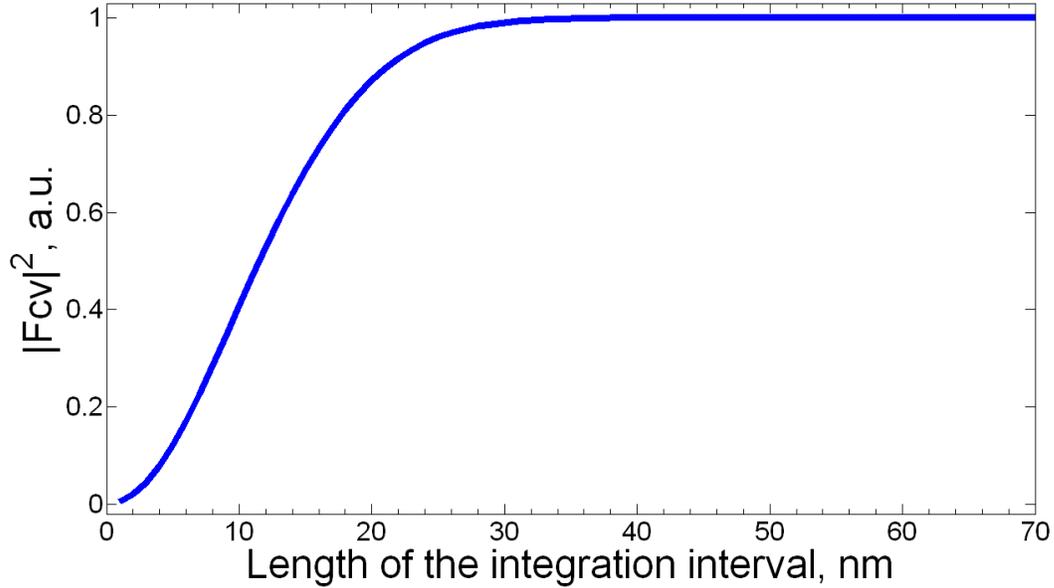

Fig. S10. Optimized, normalized absolute square of the overlap integral (2) as a function of the integration interval ($a - b$). The value of $|F_{cv}|^2$ has been maximized for each value of the integration interval as function of the lower boundary $a$, and the resulting curve has been normalized to its maximum value for the sake of clarity of presentation.

The absolute square of the overlap integral reaches 99 % of its maximum at 30 nm integration width, as shown in Fig. S10. Thus, only 30 nm of the 120 nm wide triangular quantum well is effectively contributing to a high radiative recombination rate. Hence, it is natural to adopt this thickness as the length of the gain medium per triangular quantum well, $L_{QW}$, in the GaAsSb-based superlattice. The threshold material gain required for lasing is then given by:[14]

$$g_{th,mat} = \frac{1}{2\Gamma_{core}L_{QW}n_{QW/SL}n_{SL}} \ln\left(\frac{1}{R_1 R_2}\right) \tag{3}$$

Here $\Gamma_{core}$ is the optical mode confinement factor (which was calculated to be 1.12 for the HE11$_b$ mode from FDTD simulations), $n_{QW/SL} = 10$ is the number of quantum wells per superlattice, and $n_{SL}=5$ is the number of superlattices excited with the employed spot size of around 8 µm. $R_1 = R_2 = 0.5$ is the reflectivity of the NW end facets at a wavelength of 950 nm obtained by FDTD calculations. Due to the fact that the lasing wavelength is below the band gap of the GaAs NW core material, absorption losses should be negligible and have therefore been neglected in (3). The resulting threshold material gain is from this estimated to be ~ $4.1 \times 10^3$ cm$^{-1}$. It is worth mentioning that the fact that a single quantum well was used for the calculation does not affect the end result as the modifications caused by the electronic coupling with the neighboring quantum wells occur in an area where the hole wave function is already practically zero.

# VIII. Optical pumping experiments on NW superlattices with different Sb compositions

To investigate the dependence of lasing threshold at room temperature on the Sb content in the GaAsSb superlattice, excitation power-dependent emission spectra were recorded, and spectra from six NW lasers with maximum Sb contents of ~ 1 at. % (sample A), ~ 5 at. % (sample B), and ~ 8 at. % (sample C), are shown in Fig. S11 (a) & (b), (c) & (d), and (e) & (f), respectively. As is derived from the spectra in Fig. S11, the lasing threshold for samples B and C are around 5 kW cm$^{-2}$ to 7 kW cm$^{-2}$, which is much lower than the lasing threshold of ~ 12 kW cm$^{-2}$ for sample A. This indicates that deeper GaAsSb quantum wells (higher Sb content) is beneficiary to reduce the lasing threshold.[15]

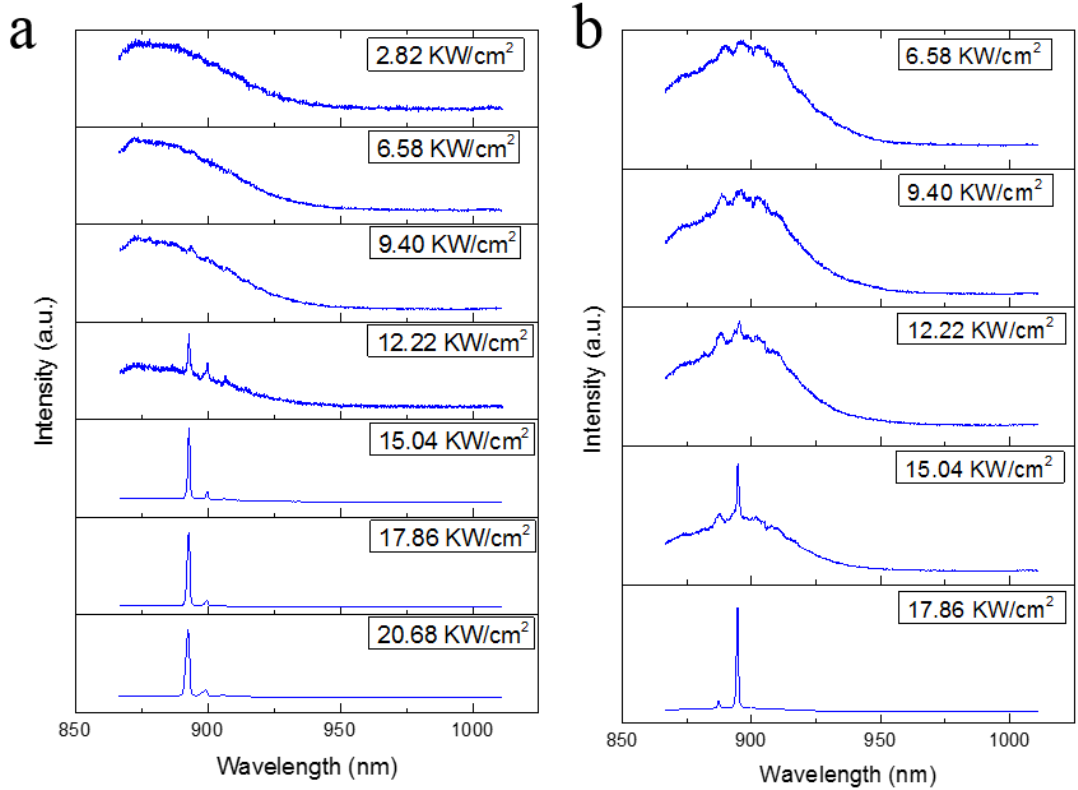

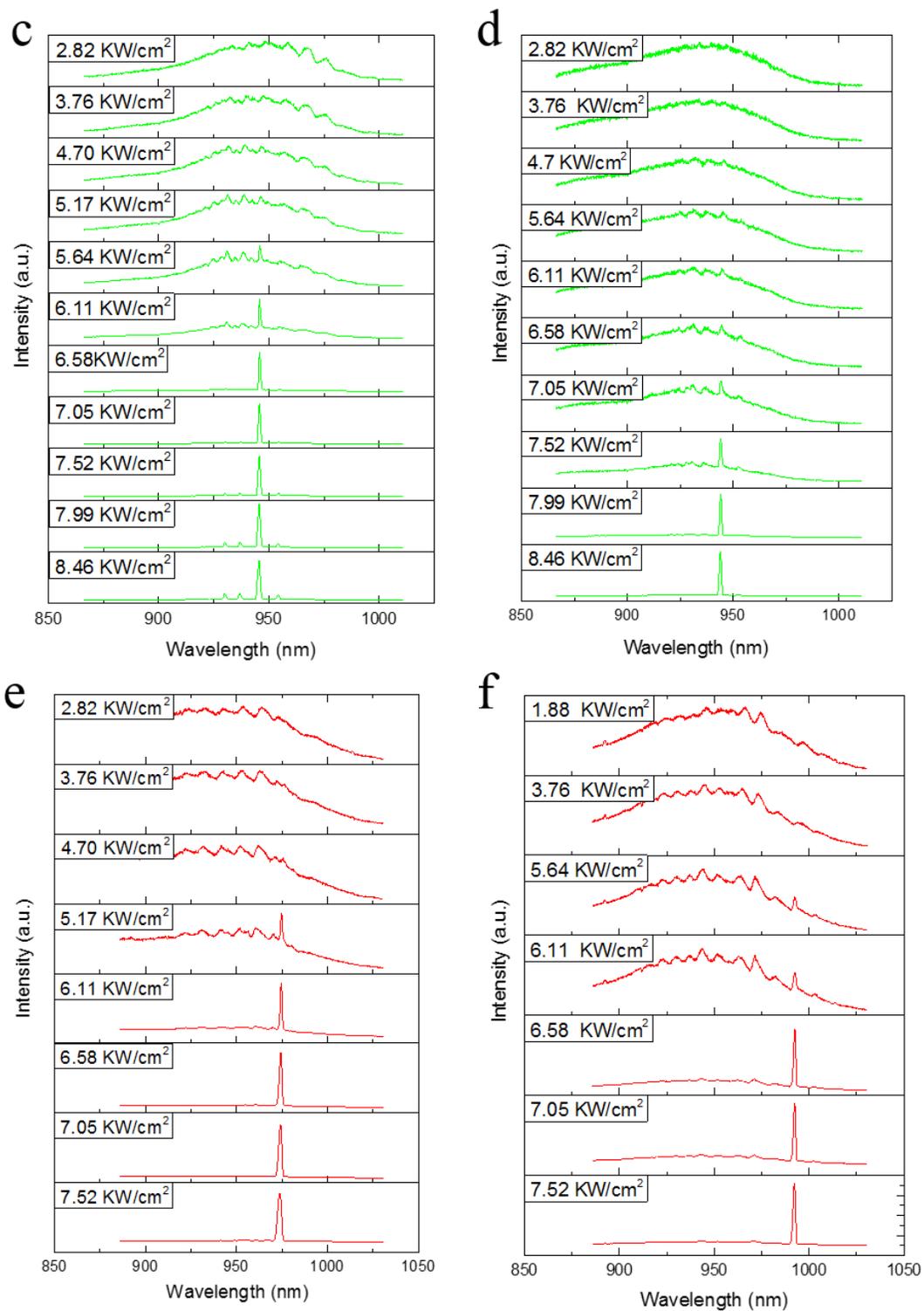

Fig. S11. Room temperature optical excitation power-dependent emission spectra for samples A, B and C. (**a, b**), (**c, d**) and (**e, f**) are from samples A, B and C with maximum Sb contents of ~ 1 at. %, ~ 5 at. %, and ~ 8 at. %, respectively.